\documentclass[a4paper]{article}
\usepackage{INTERSPEECH2014,amssymb,amsmath,epsfig}
\pdfoutput=1

\usepackage{booktabs}
\usepackage{url}
\usepackage{subfig}
\usepackage[english]{babel}
\usepackage{balance}

\title{Acoustic Gait-based Person Identification using Hidden Markov Models}
\makeatletter
\def\name#1{\gdef\@name{#1\\}}
\makeatother \name{{\em J\"urgen T. Geiger, Maximilian Knei\ss l, Bj\"orn Schuller$^1$ and Gerhard Rigoll}}

\address{Institute for Human-Machine Communication, Technische Universit\"at M\"unchen, Munich, Germany \\
 $^1$also with the Department of Computing, Imperial College London, London, U.K. \\
{\small \tt geiger@tum.de}}

\begin{document}
\ninept
\maketitle
\begin{abstract}
We present a system for identifying humans by their walking sounds.
This problem is also known as acoustic gait recognition.
The goal of the system is to analyse sounds emitted by walking persons (mostly the step sounds) and identify those persons.
These sounds are characterised by the gait pattern and are influenced by the movements of the arms and legs, 
but also depend on the type of shoe.
We extract cepstral features from the recorded audio signals and use hidden Markov models for dynamic classification.
A cyclic model topology is employed to represent individual gait cycles.
This topology allows to model and detect individual steps, leading to very promising identification rates.
For experimental validation, we use the publicly available TUM GAID database, which is a large gait recognition database containing 3\,050 recordings of 305 subjects in three variations.
In the best setup, an identification rate of 65.5\,\% is achieved out of 155 subjects.
This is a relative improvement of almost 30\,\% compared to our previous work, which used various audio features and support vector machines.
\end{abstract}

\section{Introduction}
\label{sec:intro}

Recognising people by the way they walk (also known as gait recognition or gait-based person identification) is a relatively new field of research.
Most of previously studied methods work in the visual domain, where this topic is an active field of research since the last decade~\cite{lee2002gait}.
However, acoustic information can also be used for gait recognition.
Even though the focus on this modality has so far been significantly less, results are promising.
While in the visual domain, identification systems can rely on analysing the silhouette~\cite{wang2003silhouette},
the task is much more difficult for systems working only with audio information.
The relevant information which can be exploited by such systems consists not only of the sounds of the steps,
but also adjacent sounds produced by the clothes of moving arms and legs.
These sounds are influenced by the gait pattern of the walking person,
making them suitable to be used for person identification.
Furthermore, the sounds produced during walking are highly dependent on factors such as the floor type, type of shoes and clothes.
%

In a user study~\cite{makela2003use}, the potential of humans to recognise others by their walking sounds was evaluated.
After a training phase, twelve subjects were able to identify their co-workers by their walking sounds with an accuracy of 66\,\%.
This result shows that sounds produced by walking persons convey characteristic information about the subject and can thus be used for person identification.

Potential applications of gait-based person identification using audio information are smart homes for ambient assisted living, indoor surveillance scenarios, or access control systems.
Such an audio-based system can be used to enhance visual surveillance and facilitate multimodal approaches.
As compared to video-based person identification, acoustic systems will also work in the darkness, require less expensive hardware and often lower sensor density and are less obtrusive.
Acoustic gait-based person identification is also known as \textit{acoustic gait recognition}.

\vspace{-0.1cm}
\subsection{Contribution}

The contribution of this paper is a system for acoustic gait-based person identification that is based on hidden Markov models (HMMs).
To our knowledge, this is the first time that HMMs are applied for this task.
We use Mel-frequency cepstral coefficients (MFCCs) as audio features and HMMs with a cyclic topology for dynamic classification, in order to model the dynamics of gait patterns.
With the cyclic topology, one pass through the model corresponds to a half gait cycle containing one step.
Thus, the system is capable of detecting the individual steps in a recording and using them for person identification. 
Experiments are conducted using the TUM GAID corpus, which contains 3\,050 recordings of 305 subjects in three walking variations in a realistic setup.
The recognition system is trained with normal walking style recordings and evaluated on other recordings of normal walking style as well as variations including a backpack and shoe covers.
Our experimental results show that the developed system is capable of achieving excellent recognition rates compared to previous work.


\vspace{-0.1cm}
\subsection{Related Work}

The most-widespread approach for video-based gait recognition is the Gait Energy Image (GEI)~\cite{han2006}, which is a simple silhouette-based approach.
It can be combined with face recognition~\cite{hofmann2012combined} or with depth information~\cite{hofmann2012gait}.
Furthermore, model-based approaches have been proposed for visual gait recognition~\cite{yam2004}.
Besides using video or audio information, other methods to identify walking persons include using acoustic Doppler sonar~\cite{kalgaonkar2007acoustic} or pressure sensors in the floor~\cite{yun2003user}.

Using audio information for the task of gait-based person identification is a relatively new research field.
%
In~\cite{she2004framework}, footstep sounds were detected in a corpus of various environmental sounds.
A system for person identification using footstep detection was introduced in~\cite{shoji2004personal}.
The system was tested with a database of five persons.
This work was extended in~\cite{itai2006footstep} by adding psychoacoustic features such as loudness, sharpness, fluctuation strength and roughness.
Finally, in~\cite{itai2008footstep}, dynamic time warping was used for classification and the database was extended to contain ten persons. 
The system achieves almost 100\,\% perfect classification rates (using ten persons).
However, the task is simplified by reducing it to classification of pre-segmented footsteps.
%
A similar task is addressed in the recently published study by Altaf et al.~\cite{Altaf13-PIU}.
There, a database of segmented footstep sounds from ten persons is used.
Instead of extracting spectral features,  
the shape and properties of a footstep sound are examined in a temporal energy domain.
As a result, an identification accuracy above 90\,\% is achieved by using a large number of footsteps during testing.
When using only three consecutive footsteps, which is more comparable to our work, an accuracy of 45\,\% is obtained.
Other studies on acoustic gait-based person identification were presented in ~\cite{deCarvalho2010identification,alpert2010acoustic}.
The weakness of all previous studies about acoustic gait-based person identification that are mentioned here is the fact that only small databases (mostly no more than ten subjects) that are overly prototypical have been employed.
In addition, very often, classification is performed using pre-segmented footsteps.
In our previous work~\cite{13hof1}, we investigated the potential of spectral, cepstral and energy-related audio features in combination with support vector machines (SVM) for acoustic gait-based person identification.
This work was continued in~\cite{Geiger13GBP}, where a feature analysis method was used to select relevant audio features.
In~\cite{11wen2}, we had also employed cyclic HMMs, for animal sound classification.
The cyclic model topology proved to be efficient to model the repetitive structure of these sounds.

The remainder of this paper is structured as follows:
In Section~\ref{sec:database}, we introduce the TUM GAID database which is used in the experiments.
The employed system is described in Section~\ref{sec:system}, followed by the experimental setup and results in Section~\ref{sec:experiments}.
Some concluding remarks are given in Section~\ref{sec:conclusions}.

\section{The TUM GAID Database}
\label{sec:database}

\begin{figure}[htb]
\vspace{-0.3cm}
\center{
  \subfloat[Normal recording]{\includegraphics[width=0.17\linewidth]{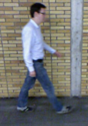}}\hspace{0.1cm}
  \subfloat[Backpack recording]{\includegraphics[width=0.17\linewidth]{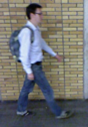}}\hspace{0.1cm}
  \subfloat[Shoe cover recording]{\includegraphics[width=0.17\linewidth]{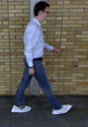}}
}

  \caption{Screenshots of three recordings in the TUM GAID database}
	\label{fig:screenshots}
\vspace{-0.2cm}
\end{figure}

For our experiments, we use our freely available\footnote{\url{www.mmk.ei.tum.de/tumgaid}} TUM Gait from Audio, Image and Depth (GAID) database~\cite{13hof1}.
The motivation behind the TUM GAID database is to foster research in multimodal gait recognition.
Therefore, data was recorded with an RGB-D sensor, as well as with a four-channel microphone array.
Thus, a typical colour video stream, a depth stream and an audio stream are simultaneously available.
The database contains recordings of 305 subjects walking perpendicular to the recording device in a 3.5\,m wide hallway corridor with a solid floor.
In each recorded sequence, the subject walks for roughly 4\,m, typically performing between 1.5 and 2.5 gait cycles (each of them consisting of two steps).
Most of the sequences have a length of approximately 2 -- 3\,s.
Three variations are recorded for each subject: Normal walking ($\mathcal{N}$), walking with a backpack ($\mathcal{B}$), and walking with shoe covers ($\mathcal{S}$).
For each subject, all recordings of the $\mathcal{N}$ condition were recorded directly after each other.
This means that the same shoes and clothes are used, which corresponds more to a re-identification scenario.
The backpack constitutes a significant variation in gait pattern and sound, and the shoe covers pose a considerable change in acoustic condition.
Figure~\ref{fig:screenshots} shows screenshots of the three different walking conditions for one subject.
For each subject, there are six recordings of the $\mathcal{N}$ setup, and two each of the $\mathcal{B}$ and $\mathcal{S}$ setups.
This sums to a total number of 3\,050 recordings.
The metadata distribution of the database is well-balanced with a female proportion of 39\,\% and ages ranging from 18 to 55 years (average 24.8 years and standard deviation 6.3 years).
More than half of the subjects are wearing sneakers while other commonly-used types of shoes are boots and loafers.

To allow for a proper scientific evaluation and to prevent overfitting on the test data, the database is divided into a \textit{development set} and a \textit{test set}.
The two sets are person-disjunct and contain 150 and 155 subjects, respectively.
Both for the development and for the test set, the first four $\mathcal{N}$ recordings of each subject are used for the enrollment process.
The other two $\mathcal{N}$ recordings as well as the $\mathcal{B}$ and $\mathcal{S}$ recordings are used to perform the identification experiments.
This means that models are learnt only using the $\mathcal{N}$ recordings, while the $\mathcal{B}$ and $\mathcal{S}$ conditions constitute previously unseen variations during the identification experiments and will therefore deteriorate the identification performance.
The partition of the database is shown in Table~\ref{Tab:databaseid}.
\begin{table}[t]
\centering
\caption{Partition of the TUM GAID database}
\vspace{-0.2cm}
\begin{tabular}{lcc}
& \textbf{Development} & \textbf{Test} \\
& (150 subj.) &(155 subj.)\\
\midrule
$\mathcal{N}$1 -- $\mathcal{N}$4 & Enrollment &  Enrollment\\
$\mathcal{N}$5 -- $\mathcal{N}$6 & Identification  &  Identification\\
$\mathcal{B}$1 -- $\mathcal{B}$2 & Identification &  Identification\\
$\mathcal{S}$1 -- $\mathcal{S}$2 & Identification &  Identification\\
\end{tabular}
\label{Tab:databaseid}
\end{table}

\section{System Description}
\label{sec:system}



We use an HMM system for classification.
Each individual subject is modelled by one HMM.
While we started with using system settings from a simple word-based speech recognition system, we modified and improved the system properties to fit to the problem of acoustic gait recognition.


\subsection{Audio Features}


In our previous work we focussed on exploring the suitability of different audio features for the problem of acoustic gait-based person identification~\cite{Geiger13GBP}.
Using SVMs for classification, we evaluated different feature sets containing MFCCs and other spectral or energy-related features.
Since SVMs are relatively robust (in contrast to HMMs) with regard to the number of employed features, we were able to improve the average identification accuracy (on the test set of the TUM GAID database) from 23.9\,\% (only MFCCs) to 28.2\,\% by adding and selecting relevant features.
In the present work, the focus is not on the front-end processing but rather on the back-end recognition system.
Therefore we keep the front-end fixed to using only MFCCs.
We use MFCC features in the standard configuration: MFCCs 0--12 including their delta and acceleration coefficients, computed every 10 $ms$ from a 25 $ms$ Hamming window, resulting in 39 features in total.
While the database provides four-channel audio recordings, we extract features from monaural recordings, which are obtained by averaging over the four channels.
In addition, we obtained slight improvements by processing the audio features with principal component analysis (PCA), without reducing the number of components.
Here, the transformations are computed only on the enrollment data, and applied on both the enrollment and identification data.

\begin{figure}
	\centering
	\subfloat{
		\includegraphics[trim = 11mm 80mm 30mm 80mm, clip, scale=0.35]{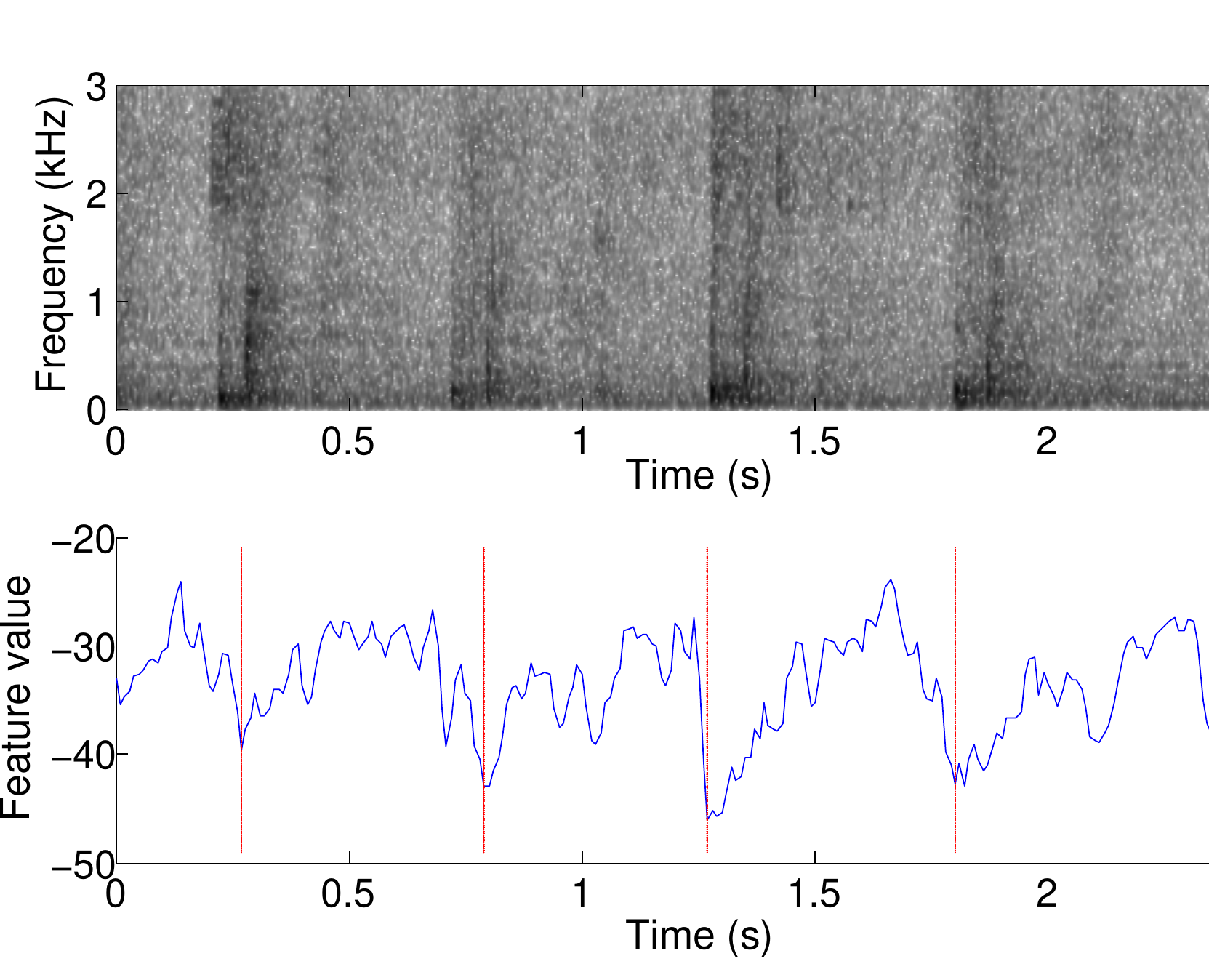}
		\label{fig:feat_spec10}
	}\\
	\vspace{-0.8cm}
	\subfloat{
		\includegraphics[trim = 11mm 80mm 30mm 80mm, clip, scale=0.35]{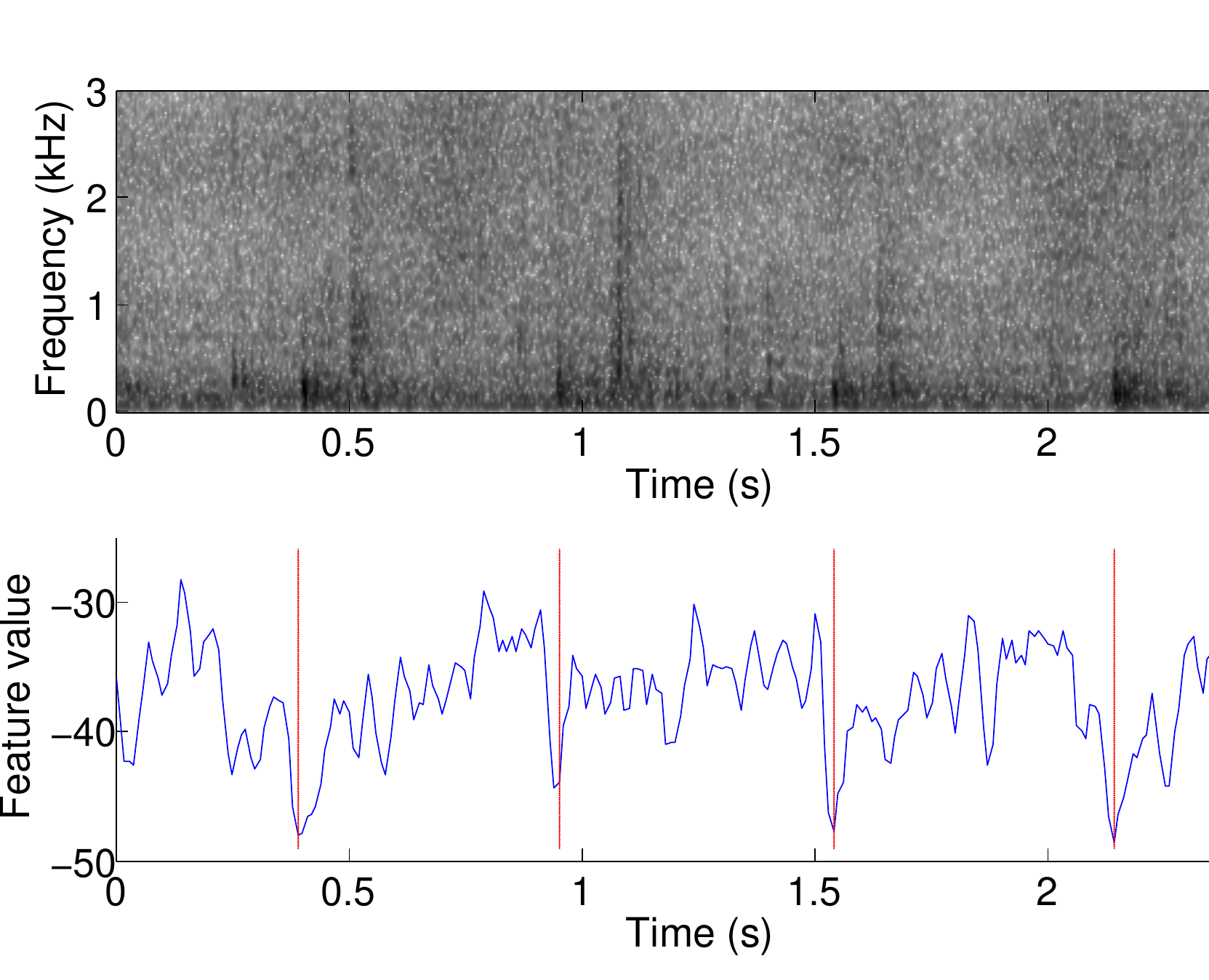}
		\label{fig:feat_spec20}
	}
	\caption{Spectrograms (top) and corresponding first MFCC coefficients (bottom), each, for a normal-type recording of two different subjects. Temporal position of footsteps is marked with a vertical line.}
	\label{fig:feat_spec}
	\vspace{-0.2cm}
\end{figure}

Figure~\ref{fig:feat_spec} shows the spectrograms and corresponding first MFCC coefficients for two exemplary recordings ($\mathcal{N}$ setup) of two different subjects.
The spectrograms reveal a considerable static background noise, which is due to the recording environment.
Several spectral peaks can be identified which correspond to the footsteps and the sounds between the steps, which are mostly made by the legs of the trousers or skirts rubbing against each other.
In the plot of the MFCCs, the temporal position of the steps are marked.
The behaviour of the MFCC features indicates that they are useful to detect the position of the steps and to distinguish between different persons.

\subsection{HMM System}
\label{ssec:basichmm}

Our starting point is a simple HMM system that can be compared to a whole-word recognition system (each person representing one \textit{word}) in speech recognition.
Each subject in the dataset is represented by an HMM.
The models are equipped with a linear left-right topology.
With such a model topology, the HMM has to pass through all of its states sequentially without skipping a state.
Before introducing an appropriate step modelling method, which will be described in the next subsection,
we apply an approach where each recording containing several steps is modelled by one pass through an HMM.
As a result, rather large numbers of states (generally more than ten) are required to be able to model the dynamic sequence of sounds during walking.

In a standard HMM system, the observations are modelled with a mixture of Gaussians.
However, our first experiments showed that the best results are obtained by using HMMs with a single Gaussian state model,
as the amount of training data is very small and hence probably not sufficient to train a more fine-grained distribution of the features.
Another reason could be that a higher number of components leads to overfitting, modelling also the noise in the recordings.

During decoding, a grammar controls the possible recognition output.
Our most simple employed grammar follows the basic HMM system setup where exactly one pass through a model is allowed for each recording.
A multi-step grammar is then introduced to let the system automatically segment the recording:
Any number of repetitions of the same model (subject) is allowed.
In order to train the HMMs to model the separate steps, an approach using a cyclic HMM topology is employed as described in the following.

\subsection{Step Modelling}
\label{sec:stepmodeling}

To be able to model the individual steps in each recording, we use cyclic HMMs.
In our basic HMM system, each recording (containing several gait cycles) is modelled by one pass through the HMM.
The strategy of representing each gait cycle separately by one pass through all states of the HMM is better suited to model the observations.
We consider the two halves of each gait cycle to be equivalent (although in fact, there is a person-dependent assymetry~\cite{nixon2006human}),
and therefore the system is designed to model half a gait cycle (containing one step) by each HMM.
In this way, one pass through the HMM models the sounds of one step and adjacent sounds (produced by the moving arms and legs). 
This method of step modelling is implemented in the system configuraton and training in the following way:
The state transition matrix of each HMM has a left-right topology, and jumps from the last state to the first state are allowed.
Models are trained with embedded re-estimation, where the number of steps is known (as determined by simple video processing methods).
As a result, the position of the steps in the training data is automatically estimated during model training.
Together with the introduced multi-step decoding grammar, the developed system is then capable of detecting, segmenting and recognising the steps occurring in the recordings.

\section{Experiments}
\label{sec:experiments}

Experiments are performed with the TUM GAID database that was described in Section~\ref{sec:database},
using the development set for system design and tuning.
Finally, we use the test set to evaluate our best system configuration.
For all systems evaluated using the development set, 15 HMM states appeared to be the optimal configuration. 
In addition, the best results were obtained with six training iterations.
For each system setup, we report experimental results (identification accuracy) separately for the three different recording conditions (normal, backpack, shoe covers).
In addition, the average accuracy over these three conditions is included.




\vspace{-0.2cm}
\subsection{Development set}

Table~\ref{tab:resultsdev} shows the results on the development set for different system configurations.
\begin{table}[t!]
\begin{center}
\caption{\em Development set (150 subjects) evaluation of different audio features, for the normal ($\mathcal{N}$), backpack ($\mathcal{B}$) and shoe cover ($\mathcal{S}$) recording conditions.}
\begin{tabular}{c|ccc|c}
  & \multicolumn{3}{|c|}{\textbf{Condition}} & \\
{\textbf{Accuracy [\%]}} & $\mathcal{N}$ & $\mathcal{B}$ & $\mathcal{S}$ & \bf average \\
\toprule
basic HMM & 53.3 & 30.7 & 7.0 & 28.2 \\ 
+ multi-step decoding & 56.3 & 31.3 & 7.3 & 31.6 \\ 
+ PCA & 57.7 & 34.3 & 9.7 & 33.9 \\ 
+ step modelling & 69.7 & 44.7 & 9.3 & 41.2 \\ 
\end{tabular}
\label{tab:resultsdev}
\end{center}
\vspace{-0.2cm}
\end{table}
The basic HMM system without explicit modelling of separate steps (cf. Section~\ref{ssec:basichmm}) is the first evaluated system.
In the normal recording condition, slightly more than half of the testing samples are classified correctly.
Averaging over the three different conditions, an accuraccy of 28.2\,\% is obtained, which serves as a baseline for further experiments.
The first step towards the improved recognition system is the introduction of a decoding grammar which allows to recognise multiple sequential instances of the same subject in the recordings.
This modification improves the average accuracy to 31.6\,\% (mostly due to improvements in the $\mathcal{N}$ setup).
Applying PCA to the features improves the accuracy for all three recording conditions.
Training the system to model each step by one pass through an HMM (cf. Section~\ref{sec:stepmodeling}) leads to the largest improvement in accuracy.
In the normal walking condition, more than two thirds of the samples are now identified correctly.
The accuracy in the backpack walking condition is also greatly improved, whereas the performance in the shoe cover condition remains largely unaffected.
While the improvements obtained with the multi-step grammar and PCA are not significant, improved step modelling leads to a significant improvement in the $\mathcal{N}$ and $\mathcal{B}$ conditions and for the average accuracy (evaluated with a one-tailed t-test with a significance level of $\alpha=0.05$).

With a simple analysis we examined the system's ability to correctly detect the individual steps.
To this end, we use the best-performing developed system (row four in Table~\ref{tab:resultsdev}).
For the test samples of the normal walking conditions, we observe the number of steps detected by the system.
The average number of steps in these test recordings is 5.3, while the system predicts 4.3 steps, on average.
For correctly identified \textit{subjects}, the average number of predicted \textit{steps} is 5.0, while for incorrectly identified subjects it is 3.5.
This shows that when the subjects are identified correctly, the step segmentation works very well.

\vspace{-0.1cm}
\subsection{Test set}

In Table~\ref{tab:resultstest}, we show the results on the test set, for our baseline system and the best system configuration.
For comparison, we include our previously published results on the same dataset.
\begin{table}[t!]
\begin{center}
\caption{\em Test set (155 subjects) evaluation of our system compared to our previously published results, for the normal ($\mathcal{N}$), backpack ($\mathcal{B}$) and shoe cover ($\mathcal{S}$) recording conditions.}
\begin{tabular}{c|ccc|c}
  & \multicolumn{3}{|c|}{\textbf{Condition}} & \\
{\textbf{Accuracy [\%]}} & $\mathcal{N}$ & $\mathcal{B}$ & $\mathcal{S}$ & \bf average \\
\toprule
video (GEI) \cite{13hof1} & 99.4 & 27.1 & 52.6 & 59.7 \\
baseline SVM \cite{13hof1} & 44.5 & 27.4 & 4.8 & 25.6 \\
SVM + feat. sel. \cite{Geiger13GBP} & 51.9 & 28.4 & 4.2 & 28.2 \\
\midrule
basic HMM & 41.0 & 24.2 & 7.1 & 24.1 \\
improved HMM & 65.5 & 36.5 & 9.0 & 37.0 \\ 
\end{tabular}
\label{tab:resultstest}
\end{center}
\vspace{-0.2cm}
\end{table}
The first row shows results of a state-of-the-art gait recognition method working with video data, namely the GEI~\cite{13hof1}.
This method achieves almost perfect results in the normal walking condition, while especially the backpack and also the shoe variation constitute a real difficulty for the system (59.7\,\% on average).
However, these results have to be interpreted carefully, since the GEI utilises mainly the appearance (the silhouette of a person) and not the behaviour (the gait pattern).
Using a large set of different audio features (1\,625 static features per recording) and SVMs for classification (second row) was our first audio-domain baseline system~\cite{13hof1}.
Naturally, the addressed task is much more difficult when dealing only with audio data (average accuracy 25.6\,\%).
However, this system can compete with the GEI in the backpack recording variation.
In~\cite{Geiger13GBP}, we improved the SVM system by employing a feature-selection technique to chose relevant features for the task,
obtaining an average identification accuracy of 28.2\,\%.
Now, with our basic HMM setup, the resulting accuracy of 24.1\,\% is comparable to the baseline SVM system.
The methods introduced in this work (primarily modelling each step separately during model training and decoding)
are able to bring a large improvement, reaching 37.0\,\%.
In the $\mathcal{N}$ and $\mathcal{B}$ recording conditions, the accuracy is improved significantly, by more than one third.
The accuracy of the video-processing method (GEI) in the backpack recording condition is surpassed by 26\,\% relatively.
Compared to the previous best-performing audio system (the SVM system including feature selection) the average accuracy is improved by 24\,\%, relatively (significant in all recording conditions).

\vspace{-0.2cm}

\section{Conclusions}
\label{sec:conclusions}

We developed a model-based system for recognising people from walking sounds.
The system uses HMMs in a cyclic topology to automatically segment the recordings according to separate steps.
Experiments were conducted using the TUM GAID database containing recordings of 305 subjects (150 in the development set and 155 in the test set) in three different recording conditions:
normal walking, walking with a backpack, and walking with shoe covers.
The results show that a basic HMM system (without explicit modelling of separate steps) achieves a similar performance in comparison to the SVM system presented in our previous work.
Improving the system with the methods introduced in this work results in large performance gains in identification accuracy.
With this system, each half gait cycle is modelled by one pass through a cyclic HMM.
This covers the sound of one step and adjacent sounds, which are mainly produced by moving arms and legs.
Thus, it is clear that the backpack or shoe cover variation influence the identification performance in a negative way.
However, when identification experiments are carried out with the same walking style and shoe type as the model was trained with (normal walking condition), almost two thirds of the subjects are identified
correctly from the test set containing 155 individuals.


Given the challenging but application-friendly enrollment of only four examples per walking subject
and in order to improve the robustness of the system, adopting approaches 
from speaker recognition like creation of models through adaption from a background model~\cite{reynolds2000speaker} could be a promising strategy in the future.
Furthermore, we will work on improving the system's robustness to variations.
This includes better coping with the backpack and shoe cover recording conditions.
In addition, the TUM GAID database contains a set of subjects with recordings made on two different dates in time (with three months in between).
Therewith, the influence of changing types of shoes and clothes as well as possibly higher variation of the walking style on the system performance can be evaluated.
In order to improve the system in this direction, we want to test approaches to address session variability known from speaker recognition (such as joint factor analysis~\cite{kenny2007joint}) as well as methods for model adaptation or feature transformation adopted from speech recognition systems.

\vfill\pagebreak

\balance
\bibliographystyle{IEEEtran}
\bibliography{tumkinectgait}

\end{document}